\documentclass[11pt]{article}
\usepackage{fa2023}
\usepackage{amsmath}
\usepackage{cite}
\usepackage{url}
\usepackage{graphicx}
\usepackage{color}
\usepackage{siunitx}
\usepackage[utf8]{inputenc}
\usepackage{babel}
\usepackage{amsfonts,mathtools}
\usepackage{dblfloatfix}% for two-column figure
\usepackage{pgfplots}
\usepackage{tikz}
\pgfplotsset{compat=newest}
\pgfplotsset{plot coordinates/math parser=false} 
\usepgfplotslibrary{patchplots}
\usepackage[subtle,title=normal,sections=normal,margins=normal,bibliography=normal]{savetrees}
\newcounter{MYtempeqncnt}
\newcommand{\vectorY}{\mathbf{y}}
\newcommand{\vectorXj}{\mathbf{x}_{j}}
\newcommand{\vectorXdDP}{\mathbf{x}_{d}^{\mathrm{DP}}}
\newcommand{\vectorXdRev}{\mathbf{x}_{d}^{\mathrm{R}}}

\newcommand{\vectorXdom}{\mathbf{x}_{d}}

\newcommand{\vectorN}{\mathbf{n}}
\newcommand{\vectorU}{\mathbf{u}_{d}}
\newcommand{\RTFvectorSpeechDom}{\mathbf{g}_{d}}
\newcommand{\RTFvectorHeadSpeechDom}{\mathbf{g}_{\mathrm{H}_{d}}}
\newcommand{\RTFvectorSpeechDomherm}{\mathbf{g}_{d}^{H}}
\newcommand{\RTFvectorHat}{\hat{\mathbf{g}}_{\mathrm{H}_{d}}}
\newcommand{\RTFvectorHatKL}{\hat{\mathbf{g}}_{\mathrm{H}_{d}}\left(k,l\right)}
\newcommand{\RTFvectorExtendedHatKL}{\hat{\mathbf{g}}_{d}\left(k,l\right)}
\newcommand{\RTFvectorBAR}{\bar{\mathbf{g}}\left(k,\theta_{i}\right)}

\newcommand{\selectionMatrixHA}{\mathbf{E}_{\rm{H}}}
\newcommand{\selectionVector}{\mathbf{e}}
\newcommand{\phiY}{\boldsymbol{\Phi}_{\mathrm{y}}}
\newcommand{\phiXd}{\boldsymbol{\Phi}_{\mathrm{x}_{d}}^{\rm{DP}}}
\newcommand{\phiUd}{\boldsymbol{\Phi}_{\mathrm{u}}}
\newcommand{\psdSpeechd}{\Phi_{X_{d}}^{\mathrm{DP}}}
\newcommand{\phiYHat}{\hat{\boldsymbol{\Phi}}_{\mathrm{y}}}

\newcommand{\phiUHat}{\hat{\boldsymbol{\Phi}}_{\mathrm{u}}}

\DeclareMathOperator{\sinc}{sinc}
\DeclarePairedDelimiter\abs{\lvert}{\rvert}%
\DeclarePairedDelimiter\norm{\lVert}{\rVert}%
\DeclarePairedDelimiter\curlyBrace\{\}
\newcommand{\myExpectation}[1]{\operatorname{\mathcal{E}}\curlyBrace*{#1}}
\newcommand{\myPrincipalEigenvec}[1]{\operatorname{\mathcal{P}}\curlyBrace*{#1}}

\title{Exploiting an External Microphone for Binaural \\ RTF-Vector-Based Direction of Arrival Estimation \\ for Multiple Speakers}

\multauthor
{Daniel Fejgin$^{*}$ \hspace{1cm} Simon Doclo \hspace{1cm}} 
{University of Oldenburg, Department of Medical Physics and Acoustics\\
and Cluster of Excellence Hearing4all, Oldenburg, Germany
\correspondingauthor{daniel.fejgin@uol.de}{Daniel Fejgin and Simon Doclo.}}

\begin{document}

\maketitle
\begin{abstract}
\let\thefootnote\relax\footnotetext{\\This work was funded by the Deutsche Forschungsgemeinschaft (DFG, German Research Foundation) under Germany’s Excellence Strategy - EXC 2177/1 - Project ID 390895286 and Project ID 352015383 - SFB 1330 B2.}In hearing aid applications, an important objective is to accurately estimate the direction of arrival (DOA) of multiple speakers in noisy and reverberant environments. Recently, we proposed a binaural DOA estimation method, where the DOAs of the speakers are estimated by selecting the directions for which the so-called Hermitian angle spectrum between the estimated relative transfer function (RTF) vector and a database of prototype anechoic RTF vectors is maximized. The RTF vector is estimated using the covariance whitening (CW) method, which requires a computationally complex generalized eigenvalue decomposition. The spatial spectrum is obtained by only considering frequencies where it is likely that one speaker dominates over the other speakers, noise and reverberation. In this contribution, we exploit the availability of an external microphone that is spatially separated from the hearing aid microphones and consider a low-complexity RTF vector estimation method that assumes a low spatial coherence between the undesired components in the external microphone and the hearing aid microphones. Using recordings of two speakers and diffuse-like babble noise in acoustic environments with mild reverberation and low signal-to-noise ratio, simulation results show that the proposed method yields a comparable DOA estimation performance as the CW method at a lower computational complexity. 
\end{abstract}
\section{Introduction}\label{sec:introduction}
In speech communication applications such as hearing aids, methods for estimating the direction of arrival (DOA) of multiple speakers are often required. To solve this estimation task, (deep) learning-based and model-based methods are continuously developed and advanced \cite{Huang2008,Grumiaux2022}. However, only few methods exploit the availability of external mobile devices equipped with microphones \cite{Farmani2018,Kowalk2022,Fejgin2021,Fejgin2023}, although wirelessly linking hearing aids to these devices has become increasingly popular \cite{Mecklenburger2016}. 

Recently, we proposed relative-transfer-function (RTF) vector-based DOA estimation methods for a single speaker in \cite{Fejgin2021,Fejgin2023}, without relying on the external microphone to be close to the target speaker and capturing only little noise or reverberation as in \cite{Farmani2018,Kowalk2022}. We estimated the DOA as the direction that maximized the similarity between the estimated RTF vector and a database of prototype anechoic RTF vectors for different directions in terms of a frequency-averaged distance function. 

However, the methods in \cite{Farmani2018,Fejgin2021,Kowalk2022,Fejgin2023} considered only a single speaker. To address DOA estimation for multiple speakers, we introduced the so-called frequency-averaged Hermitian angle spectrum from which the DOAs were estimated as the directions corresponding to the peaks of this spatial spectrum (throughout the paper, we refer to a direction-dependent similarity score as a spatial spectrum) \cite{Fejgin2022}. Opposed to \cite{Fejgin2021,Fejgin2023}, the spatial spectrum was constructed from time-frequency (TF) bins where one speaker was assumed to be dominant over all other speakers, noise, and reverberation, solely.

Estimation of the RTF vector of a speaker from noisy microphone signals can be accomplished using, e.g., the state-of-the-art covariance whitening (CW) method \cite{Markovich2009} or the spatial coherence (SC) method \cite{Goessling2018}. Despite the effectiveness of the CW method and the possibility to apply the method using only the head-mounted microphone signals or all available signals, such a computationally expensive method (due to the inherent generalized eigenvalue decomposition) is less desirable than methods with a lower computation complexity for resource-constrained applications like hearing aids. Opposed to the CW method, the SC method requires an external microphone but does not perform expensive matrix decompositions. The SC method relies on the assumption of a low spatial coherence between the undesired component in one of the microphone signals and the undesired components in the remaining microphone signals. As shown in \cite{Goessling2018}, this assumption holds quite well, for example, when the distance between the external microphone and the head-mounted microphones is large enough and the undesired component is spatially diffuse-like.

In this paper, we propose to construct the frequency-averaged Hermitian angle spectrum for DOA estimation for multiple speakers using the computationally inexpensive SC method. We compare the DOA estimation accuracy when estimating the RTF vector using the SC method or the CW method in a reverberant acoustic scenario with diffuse-like babble noise. Experimental results show for multiple positions of the external microphone that estimating the RTF vector with the SC method yields a DOA estimation accuracy that is comparable to the CW method at a lower computational complexity.

\begin{figure*}[!b]
	\hrule
	\vskip1pt
	\normalsize
	\setcounter{MYtempeqncnt}{\value{equation}}
	\setcounter{equation}{13}
	\begin{equation}
		\label{eq:SchwarzCDR}
		\scalebox{0.96}{$
			f\left(\widehat{\Gamma}_{\mathrm{y},i,j},\widetilde{\Gamma}_{\mathrm{u},i,j}\right) = \frac{\widetilde{\Gamma}_{\mathrm{u},i,j}\,\Re\{\widehat{\Gamma}_{\mathrm{y},i,j}\} -\abs{\widehat{\Gamma}_{\mathrm{y},i,j}}^{2} - \sqrt{{\widetilde{\Gamma}_{\mathrm{u},i,j}}^2\, {\Re\{\widehat{\Gamma}_{\mathrm{y},i,j}\}}^{2} - {\widetilde{\Gamma}_{\mathrm{u},i,j}}^{2}\,{\abs{\widehat{\Gamma}_{\mathrm{y},i,j}}}^{2} + {\widetilde{\Gamma}_{\mathrm{u},i,j}}^{2} - 2\, \widetilde{\Gamma}_{\mathrm{u},i,j}\, \Re\{\widehat{\Gamma}_{\mathrm{y},i,j}\} + {\abs{\widehat{\Gamma}_{\mathrm{y},i,j}}}^2}}{{\abs{\widehat{\Gamma}_{\mathrm{y},i,j}}}^{2} - 1}$}
	\end{equation}
	\setcounter{equation}{\value{MYtempeqncnt}}
\end{figure*}
\section{Signal model and notation}
We consider a binaural hearing aid setup with $M$ microphones, i.e., $M/2$ microphones on each hearing aid, and one external microphone that is spatially separated from the head-mounted microphones and can be located at an arbitrary position, i.e., $M +1$ microphones in total. We consider an acoustic scenario with $J$ si\-mul\-taneously active speakers with DOAs $\theta_{1:J}$ (in the azimuthal plane) in a noisy and reverberant environment, where $J$ is assumed to be known. In the short-time Fourier transform (STFT) domain, the $m$-th microphone signal can be written as
\begin{equation}
	Y_{m}\left(k,l\right) = \sum_{j=1}^{J}X_{m,j}\left(k,l\right) + N_{m}\left(k,l\right)\,, \label{eq:signalModel_micComponent}
\end{equation}
where $m \in \left\{1,\dots,M+1\right\}$ denotes the microphone index, $k\in\left\{1,\dotsc,K\right\}$ and $l\in\left\{1,\dotsc,L\right\}$ denote the frequency bin index and the frame index, respectively, and $X_{m,j}\left(k,l\right)$ and $N_{m}\left(k,l\right)$ denote the $j$-th speech component and the noise component in the $m$-th microphone signal, respectively. For conciseness, we will omit the frequency bin index $k$ and the frame index $l$ in the remainder of this paper wherever possible. Assuming sparsity in the STFT domain and one dominant speaker (indexed by $j=d$) per TF bin \cite{Yilmaz2004}, and stacking all microphone signals in an $\left(M+1\right)$-dimensional vector $\vectorY =\left[Y_{1},\medspace\medspace\medspace\dots\medspace\medspace\medspace, Y_{M+1}\right]^{T}$, where $\left(\cdot\right)^{T}$ denotes transposition, the vector $\vectorY$ is given by
\begin{equation}
	\vectorY = \sum_{j=1}^{J}\vectorXj + \vectorN \approx \vectorXdom + \vectorN\,,
	\label{eq:signalModel_vectorized}
\end{equation} 
with $\vectorXj$, $\vectorXdom$, and $\vectorN$ defined similarly as $\vectorY$.	

Choosing the first microphone as the reference microphone (without loss of generality) and assuming that the speech component for each (dominant) speaker can be decomposed into a direct-path component $\vectorXdDP$ and a reverberant component $\vectorXdRev$, $\vectorXdom$ can be written as
\begin{equation}
	\vectorXdom = \vectorXdDP + \vectorXdRev = \RTFvectorSpeechDom X_{1,d}^{\rm  DP} + \vectorXdRev\,,
	\label{eq:domSpeech_RTF}
\end{equation}
where 
\begin{equation}
	\RTFvectorSpeechDom = \left[1, G_{2},\dots, G_{M+1}\right]^{T}
	\label{eq:definitionRTFvec}
\end{equation}
denotes the extended $\left(M+1\right)$-dimensional direct-path RTF vector and $X_{1,d}^{\rm  DP}$ denotes the direct-path speech component of the dominant speaker in the reference microphone. The $M$-dimensional head-mounted direct-path RTF vector $\RTFvectorHeadSpeechDom$ corresponding to the head-mounted microphone signals can be extracted from $\RTFvectorSpeechDom$ as
\begin{equation}
	\RTFvectorHeadSpeechDom = \selectionMatrixHA\RTFvectorSpeechDom\,,\quad \selectionMatrixHA = \left[\mathbf{I}_{M\times M},\mathbf{0}_{M}\right]\,,\label{eq:definitionRTFvecHead}
\end{equation}
where $\selectionMatrixHA$ denotes the $\left(M\times M+1\right)$-dimensional selection matrix for the head-mounted microphone signals with $\mathbf{I}_{M\times M}$ denoting an $\left(M\times M\right)$-dimensional identity matrix and $\mathbf{0}_{M}$ denoting an $M$-dimensional vector of zeros. Both RTF vectors $\RTFvectorSpeechDom$ and $\RTFvectorHeadSpeechDom$ encode the DOA of the dominant speaker. However, the extended RTF vector $\RTFvectorSpeechDom$ depends on the (unknown) position of the external microphone, whereas the head-mounted RTF vector $\RTFvectorHeadSpeechDom$ with fixed relative positions of the head-mounted microphones (ignoring small movements of the hearing aids due to head movements) does not depend on the position of the external microphone. Hence, for DOA estimation, we will only consider the head-mounted RTF vector $\RTFvectorHeadSpeechDom$.

The noise and reverberation components are condensed into the undesired component $\vectorU = \vectorXdRev + \vectorN$ such that $\vectorY \approx \vectorXdDP + \vectorU$.

Assuming uncorrelated direct-path speech and undesired components, the covariance matrix of the noisy microphone signals can be written as
\begin{equation}
	\phiY = \mathcal{E}\left\{\vectorY\vectorY^{H}\right\} = \phiXd + \phiUd\,,
	\label{eq:definitionPhiy}
\end{equation}
with 
\begin{equation}
	\phiXd = \RTFvectorSpeechDom \RTFvectorSpeechDomherm\psdSpeechd\,, \quad \phiUd = \mathcal{E}\left\{\vectorU\vectorU^{H}\right\}\,,
\end{equation}
where $\left(\cdot\right)^{H}$ and $\mathcal{E}\left\{\cdot\right\}$ denote the complex trans\-position and expec\-tation operator, respectively. $\phiXd$ and $\phiUd$ denote the covariance matrices of the direct-path dominant speech component and undesired component, respectively, and $\psdSpeechd =\mathcal{E}\left\{\lvert X_{1,d}^{\rm  DP}\rvert^{2}\right\}$ denotes the power spectral density of the direct-path dominant speech component in the reference microphone.

\section{RTF-Vector-Based DOA Estimation}\label{sec:DOAest}
In this section, we review the RTF-vector-based DOA estimation method proposed in \cite{Fejgin2022} that is based on finding the directions corresponding to the peaks of the spatial spectrum called frequency-averaged Hermitian angle spectrum.

To estimate the DOAs $\theta_{1:J}$ of the speakers from the estimated head-mounted\footnote{As previously stated, we only consider the estimated head-mounted RTF vector $\RTFvectorHatKL$ for DOA estimation and not the extended RTF vector $\RTFvectorExtendedHatKL$ that depends both on the speaker DOA and the (unknown) position of the external microphone.} RTF vector $\RTFvectorHat\left(k,l\right)$, the estimated head-mounted RTF vector $\RTFvectorHat\left(k,l\right)$ is compared to a database of prototype anechoic RTF vectors $\RTFvectorBAR$ for several directions $\theta_{i}\,,\medspace i=1,\medspace\medspace\dotsc\medspace\medspace, I$ using the Hermitian angle \cite{Varzandeh2017} as a measure of dissimilarity, i.e.,
\begin{align}
	p\left(k,l,\theta_{i}\right) &= h\left(\hat{\mathbf{g}}_{\mathrm{H}_{d}}\left(k,l\right),\RTFvectorBAR\right)\,,\label{eq:hermitianAngleNarrow_generic}\\
	h\left(\hat{\mathbf{g}},\bar{\mathbf{g}}\right) &= \arccos\left(\frac{\abs{\bar{\mathbf{g}}^{H}\hat{\mathbf{g}}}}
	{\norm{\bar{\mathbf{g}}}_{2}\,\norm{\hat{\mathbf{g}}}_{2}}\right)\,.
\end{align} 
These prototype anechoic head-mounted RTF vectors can be obtained, e.g., via measurements using the same microphone array configuration as used during the actual source localization or using spherical diffraction models \cite{Duda1998}. 

Accounting for the disjoint activity of the speakers in the STFT domain and aiming at including only TF bins where the estimated head-mounted RTF vector $\RTFvectorHat\left(k,l\right)$ is a good estimate for the direct-path RTF vector in \eqref{eq:definitionRTFvecHead} (of one of the speakers), the narrowband spatial spectrum \eqref{eq:hermitianAngleNarrow_generic} is integrated over a set $\mathcal{K}\left(l\right)$ of selected frequency bins, where it is likely that one speaker dominates over all other speakers, noise, and reverberation \cite{Fejgin2022}, i.e.,
\begin{equation}
	P\left(l,\theta_{i}\right)=-\sum_{k\in\mathcal{K}\left(l\right)}p\left(k,l,\theta_{i}\right)\,.\label{eq:defHermitianAngleSpectrum}
\end{equation}
Based on the usage of the Hermitian angle for the construction of \eqref{eq:hermitianAngleNarrow_generic}, the spatial spectrum in \eqref{eq:defHermitianAngleSpectrum} is called the frequency-averaged Hermitian angle spectrum. The DOAs $\theta_{1:J}\left(l\right)$ are estimated by selecting the directions corresponding to the $J$ peaks of this spatial spectrum (assuming $J$ to be known).

In the context of DOA estimation, coherence-based quantities such as the coherent-to-diffuse ratio (CDR) are a common criterion for frequency subset selection \cite{Taseska2017,Brendel2018,Evers2018,Lee2020,Fejgin2022}. The usage of the CDR as a criterion for frequency subset selection can be motivated by the fact, that for higher values of the CDR at the respective TF bin it is more likely that a speaker dominates over all other speakers, noise, and reverberation at the respective TF bin. As in \cite{Fejgin2022}, the subset $\mathcal{K}\left(l\right)$ is obtained using the coherent-to-diffuse ratio (CDR) criterion \eqref{eq:CDRcriterion}, i.e.,
\begin{equation}
	\mathcal{K}\left(l\right) = \left\{k: \widehat{\mathrm{CDR}}\left(k,l\right)\geq\mathrm{CDR}_{\mathrm{thresh}}\right\}\,,\label{eq:CDRcriterion}
\end{equation}
where the CDR is estimated as
\begin{equation}
	\widehat{\mathrm{CDR}}\left(k,l\right) = f\left(\widehat{\Gamma}_{y,\mathrm{eff}}\left(k,l\right), \widetilde{\Gamma}_{u}\left(k\right)\right)\,,\label{eq:defCDR}
\end{equation}
with the CDR-functional $f$ defined in \eqref{eq:SchwarzCDR} for a single microphone pair comprising the microphones $m=i$ and $m=j$ \cite{Schwarz2015}. The arguments of the function in $\eqref{eq:SchwarzCDR}$ are the estimated coherence $\widehat{\Gamma}_{\mathrm{y},i,j}$ of the noisy signal 
\begin{equation}
	\widehat{\Gamma}_{y_{i,j}}\left(k,l\right)= \hat{\Phi}_{\mathrm{y}_{i,j}}\left(k,l\right)/\sqrt{\hat{\Phi}_{\mathrm{y}_{i,i}}\left(k,l\right)~\hat{\Phi}_{\mathrm{y}_{j,j}}\left(k,l\right)}
\end{equation}
with $\hat{\Phi}_{\mathrm{y}_{i,j}}$ denoting an estimate of the $\left(i,j\right)$-th element of the covariance matrix of the noisy microphone signals and a model $\widetilde{\Gamma}_{\mathrm{u},i,j}$ of the coherence of the undesired component. To consider more than just a single microphone pair for the estimation of the CDR, the coherence of the noisy signals between multiple microphone pairs (denoted as the microphone set $\mathcal{M}$) between the left and the right hearing aid is averaged prior to evaluating the CDR-functional in \eqref{eq:SchwarzCDR}, resulting in the binaural effective coherence \cite{Loellmann2020,Fejgin2022}, i.e.,
\begin{equation}
	\addtocounter{equation}{1}
	\widehat{\Gamma}_{y,\mathrm{eff}}\left(k,l\right) = \frac{1}{\left|\mathcal{M}\right|}\sum\limits_{i,j \in \mathcal{M}}\widehat{\Gamma}_{y_{i,j}}\left(k,l\right)\,,\label{eq:binauralEffCoh}\\
\end{equation}
Thus, the binaural effective coherence represents the average coherence between the head-mounted microphone signals. Due to the arbitrary position of the external microphone, we consider only the head-mounted microphones (with fixed relative positions) for the estimation of the binaural effective coherence $\widehat{\Gamma}_{y,\mathrm{eff}}\left(k,l\right)$. 

To model the coherence of the undesired component for the estimation of the CDR in \eqref{eq:SchwarzCDR} between the head-mounted microphone signals, head shadow effects need to be included. Assuming a diffuse sound field for both the noise and reverberation component, a modified sinc-model \cite{Lindevald1986} is employed, i.e.,
\begin{equation}
	\widetilde{\Gamma}_{u}\left(k\right) = \sinc{\left(\alpha\frac{\omega_{k}r}{c}\right)}\,\frac{1}{\sqrt{1 + \left(\beta\frac{\omega_{k}r}{c}\right)^{4}}}\,,\label{eq:modifiedSincCoherence}
\end{equation}
where $\omega_{k}$ denotes the discrete angular frequency, $r$ denotes the distance between the microphones of left and right hearing aid which is approximated as the diameter of a head, $c$ denotes the speed of sound, and $\alpha=0.5$ and $\beta=2.2$ denote empirically determined parameters of the modified sinc-model.

In this paper we compare the influence of different RTF vector estimation methods on constructing the frequency-averaged Hermitian angle spectrum in \eqref{eq:defHermitianAngleSpectrum}. In \cite{Fejgin2022} no external microphone was used and therefore the DOAs were estimated from the spatial spectrum as in \eqref{eq:spatialSpectrumIwaenc} constructed from head-mounted RTF vectors that were estimated using the CW method as in \eqref{eq:estRTF_CW}, i.e.,
\begin{equation}
	P^{\left(\mathrm{CW}\right)}\left(l,\theta_{i}\right)=-\sum_{k\in\mathcal{K}\left(l\right)}h\left(\hat{\mathbf{g}}_{\mathrm{H}_{d}}^{\mathrm{(CW)}}\left(k,l\right),\RTFvectorBAR\right)\,.\label{eq:spatialSpectrumIwaenc}
\end{equation}
In this paper, we propose to exploit the availability of the external microphone and estimate the DOAs from the spatial spectrum constructed as in \eqref{eq:spatialSpectrum_SC} constructed from head-mounted RTF vectors that are estimated using the SC method as in \eqref{eq:estRTF_SC}, i.e.,
\begin{equation}
	\boxed{P^{\left(\mathrm{SC}\right)}\left(l,\theta_{i}\right)=-\sum_{k\in\mathcal{K}\left(l\right)}h\left(\hat{\mathbf{g}}_{\mathrm{H}_{d}}^{\mathrm{(SC)}}\left(k,l\right),\RTFvectorBAR\right)}\label{eq:spatialSpectrum_SC}
\end{equation}
A summary on the covariance whitening (CW) method \cite{Markovich2009} and the spatial coherence (SC) method \cite{Goessling2018} is provided in the next section.

\section{RTF Vector Estimation}\label{sec:RTFest}
In order to estimate DOAs of multiple speakers, a frequency-averaged Hermitian angle spectrum is constructed, which assess the similarity between the estimated $M$-dimensional head-mounted RTF vector $\RTFvectorHatKL$ and a database of prototype anechoic RTF vectors for different directions. In this section, we review two RTF vector estimation methods. The computationally expensive state-of-the-art covariance whitening (CW) method \cite{Markovich2009} is summarized in Section \ref{subsec:RTFest__CW}. The computationally inexpensive spatial coherence (SC) method \cite{Goessling2018} is discussed in Section \ref{subsec:RTFest__SC}.
\subsection{Covariance whitening (CW)}\label{subsec:RTFest__CW}
To apply the CW method \cite{Markovich2009}, estimates $\phiYHat$ and $\phiUHat$ of the covariance matrices of the noisy signal and the undesired signal component are required. Based on these estimates, the head-mounted direct-path RTF vector $\RTFvectorHeadSpeechDom$ can be estimated using only the head-mounted microphone signals as
\begin{align}
	\RTFvectorHat^{\mathrm{(CW)}} &=f\left(\selectionMatrixHA\phiYHat\selectionMatrixHA^{H},\selectionMatrixHA\phiUHat\selectionMatrixHA^{H}\right)\label{eq:estRTF_CW}\,,\\
	f\left(\check{\boldsymbol{\Phi}}_{\mathrm{y}},\check{\boldsymbol{\Phi}}_{\mathrm{u}}\right)&=\frac{\check{\boldsymbol{\Phi}}_{\mathrm{u}}^{1/2}\myPrincipalEigenvec{\check{\boldsymbol{\Phi}}_{\mathrm{u}}^{-1/2}\check{\boldsymbol{\Phi}}_{\mathrm{y}}\check{\boldsymbol{\Phi}}_{\mathrm{u}}^{-H/2}}}{\check{\mathbf{e}}_{1}^{T}\check{\boldsymbol{\Phi}}_{\mathrm{u}}^{1/2}\myPrincipalEigenvec{\check{\boldsymbol{\Phi}}_{\mathrm{u}}^{-1/2}\check{\boldsymbol{\Phi}}_{\mathrm{y}}\check{\boldsymbol{\Phi}}_{\mathrm{u}}^{-H/2}}}\,,
\end{align}
where $\myPrincipalEigenvec{\cdot}$ denotes the principal eigenvector of a matrix, $\check{\boldsymbol{\Phi}}_{\mathrm{u}}^{1/2}$ denotes a square-root decomposition (e.g., Cholesky decomposition) of the $\check{M}$-dimensional matrix $\check{\boldsymbol{\Phi}}_{\mathrm{u}}$ and $\check{\mathbf{e}}_{1}=\left[1,0,\dotsc,0\right]^{T}$ denotes an $\check{M}$-dimensional selection vector. Note that $\RTFvectorHeadSpeechDom$ can be estimated likewise from the head-mounted microphone signals \textit{and} the external microphone signal together, via $\selectionMatrixHA f\left(\phiYHat,\phiUHat\right)$, differing in general from the estimate $\RTFvectorHat^{\mathrm{(CW)}}$ as in \eqref{eq:estRTF_CW}. However, based on the results of \cite{Fejgin2021} and \cite{Fejgin2023}, we will consider only the estimate as in \eqref{eq:estRTF_CW} obtained from the head-mounted microphone signals only as no significant benefit in DOA estimation performance was reported when all microphone signals were used.

\subsection{Spatial coherence (SC)}\label{subsec:RTFest__SC}
The SC method \cite{Goessling2018} requires an external microphone and relies on the assumption of a low spatial coherence between the undesired component $U_{M+1}$ in the external microphone signal and the undesired components $U_{m},\medspace m\in\left\{1,\dotsc,M\right\}$, in the head-mounted microphone signals, i.e.
\begin{equation}
	\myExpectation{U_{m}U_{M+1}^{\ast}}\approx 0\,,\quad m\in\left\{1,\medspace\medspace\medspace\dots\medspace\medspace\medspace, M\right\}\,.
\end{equation}
As shown in \cite{Goessling2018}, this assumption holds quite well, for example, when the distance between the external microphone and the head-mounted microphones is large enough and the undesired component is spatially diffuse-like. Exploiting this assumption, results in $\myExpectation{Y_{m}Y_{M+1}^{\ast}}=\myExpectation{X_{m}X_{M+1}^{\ast}},\medspace m\in\left\{1,\medspace\medspace\medspace\dots\medspace\medspace\medspace, M\right\}$, thus the RTF vector can be efficiently estimated without expensive matrix decompositions as
\begin{equation}
	\RTFvectorHat^{\mathrm{(SC)}} = \selectionMatrixHA\,\frac{\phiYHat\selectionVector_{M+1}}{\selectionVector_{1}^{T}\phiYHat\selectionVector_{M+1}}\label{eq:estRTF_SC}\,,
\end{equation}
with $\selectionVector_{m}$ denoting an $\left(M+1\right)$-dimensional selection vector selecting the $m$-th element.

\section{Experimental Results}\label{sec:Experiments}
Applying the CW and SC method for RTF vector estimation, in this section we compare the DOA estimation performance when using the SC-based frequency-averaged Hermitian angle spectrum as in \eqref{eq:spatialSpectrum_SC} against the DOA estimation performance when using the CW-based frequency-averaged Hermitian angle spectrum as in \eqref{eq:spatialSpectrumIwaenc}. We evaluate the methods with recorded signals for an acoustic scenario with two static speakers in a reverberant room with diffuse-like babble noise. The experimental setup and implementation details of the algorithms are described in Section \ref{subsec_Experiments__details}. The results in terms of localization accuracy are presented and discussed in Section \ref{subsec_Experiments__res}.

\begin{figure}
	\centerline{%
	\includegraphics[width=0.65\linewidth,trim={2cm 5cm 2cm 5cm},clip]{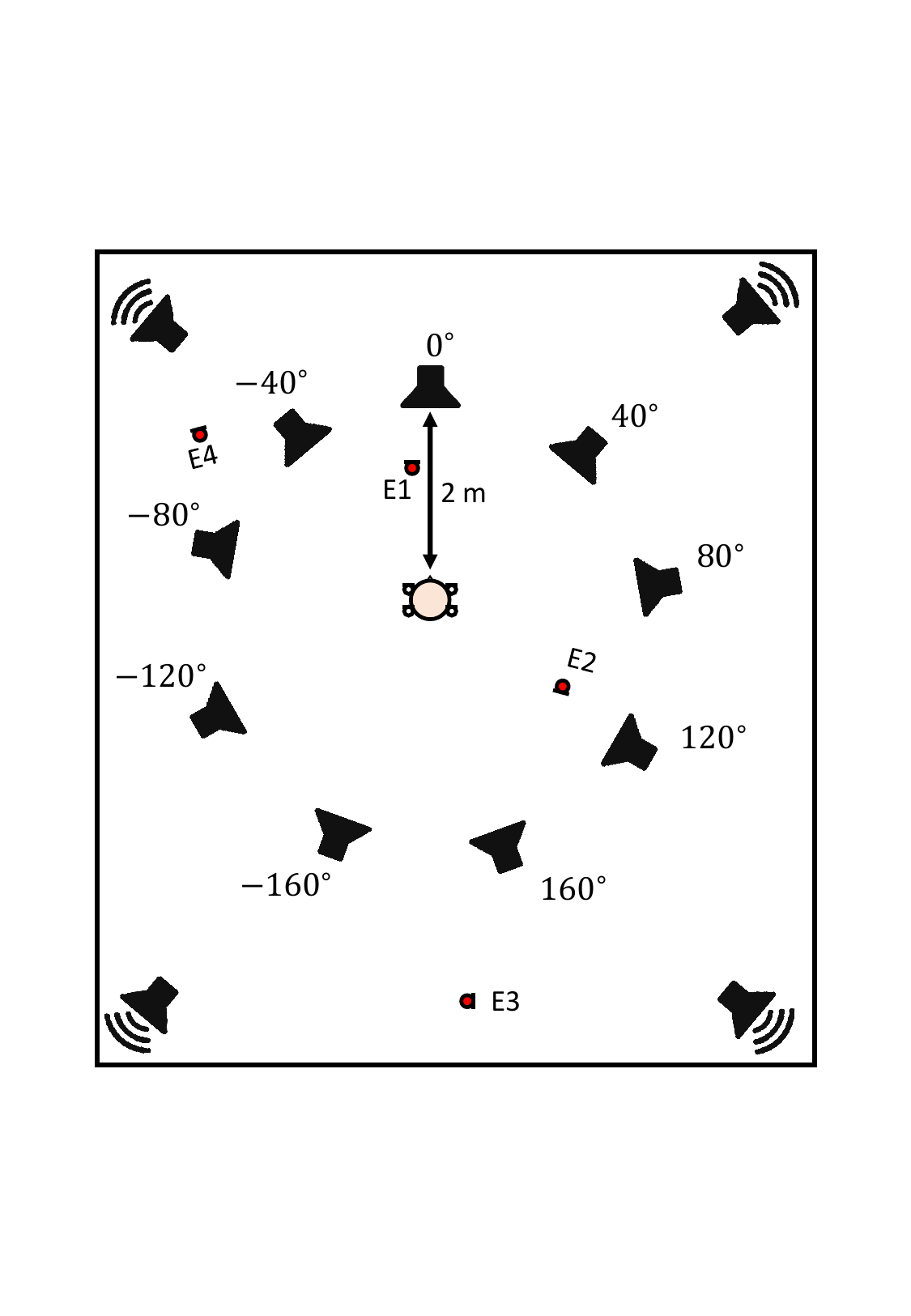}}%
	\vskip-0.4cm\caption{Experimental setup with a head-mounted binaural hearing setup and an external microphone depicted in red at four different positions (E1-E4).}\vskip-0.25cm
	\label{fig:setup}
\end{figure}

\subsection{Experimental setup and implementation details}\label{subsec_Experiments__details}
For the experiments we used signals that were recorded in a laboratory at the University of Oldenburg with dimensions of about \qty[parse-numbers=false]{7\times6\times2.7}{\cubic\meter}, where the reverberation time can be adjusted by means of absorber panels, which are mounted to the walls and the ceiling. The reverberation time was set to approximately $T_{\rm 60}\approx \qty{250}{\milli\second}$. Fig. \ref{fig:setup} depicts the experimental setup. A dummy head with a binaural hearing aid setup ($M = 4$) was placed approximately in the center of the laboratory. For this hearing aid setup a database of prototype anechoic RTF vectors is obtained from measured anechoic binaural room impulse responses \cite{Kayser2009} with an angular resolution of \qty{5}{\degree} ($I = 72$). A single external microphone was placed at four different positions (denoted as E1 - E4), which was not restricted to be close to a speaker. Two speakers from the EBU SQAM CD corpus \cite{EBU2008} (male and female, English language) were played back via loudspeakers that were located at approximately \qty{2}{\meter} distance from the dummy head. For the evaluation, all 72 pairs of DOAs of non-collocated speakers (each of the 9 DOAs in the range $\left[\qty{-160}{\degree},\qty{-120}{\degree},\dotsc,\qty{160}{\degree}\right]$) were considered. The speech signals were constantly active and had a duration of approximately \qty{5}{\second}. Diffuse-like noise was generated with four loudspeakers facing the corners of the laboratory, playing back different multi-talker recordings. The speech and noise components were recorded separately and were mixed at $\left\{\qty{-5}{\decibel},\medspace\qty{0}{\decibel},\medspace \qty{5}{\decibel}\right\}$ broadband signal-to-noise ratio (SNR) averaged over all head-mounted microphones of the hearing aid setup. All microphone signals were recorded simultaneously, hence neglecting synchronization and latency aspects.

The microphone signals were processed in the STFT-domain using a \qty{32}{\milli\second} square-root Hann window with 50 \% overlap at a sampling frequency of \qty{16}{\kilo\hertz}. The covariance matrices $\phiY$ and $\phiUd$ were estimated recursively during detected speech-and-noise and noise-only TF bins, respectively, using smoothing factors corresponding to time constants of \qty{250}{\milli\second} for $\phiYHat$ and \qty{500}{\milli\second} for $\phiUHat$, respectively. The speech-and-noise TF bins were discriminated from noise-only TF bins based on the speech presence probability \cite{Gerkmann2012}, averaged and thresholded over all head-mounted microphone signals.

We assess the DOA estimation performance by averaging the localization accuracy over the considered DOA pairs and SNRs. For the localization accuracy we average the per-frame-accuracies over all frames, where we define the per-frame accuracy as
\begin{equation}
	\mathrm{ACC}\left(l\right) = j_{\mathrm{correct}}\left(l\right)/J\,,
\end{equation}
with $j_{\mathrm{correct}}\left(l\right)$ denoting the number of speakers that are correctly localized within a range of $\pm 5^{\circ}$ in the $l$-th frame and $J=2$.

\begin{figure}
	\centering% This file was created by matlab2tikz.
\definecolor{mycolor1}{rgb}{0.00000,0.44700,0.74100}%
\definecolor{mycolor2}{rgb}{0.85000,0.32500,0.09800}%
\begin{tikzpicture}[scale=0.425]

\begin{axis}[%
width=5in,
height=5in,
at={(1.297in,1.093in)},
scale only axis,
bar shift auto,
xmin=-0.2,
xmax=10.2,
xtick={1,2,3,4,5,6,7,8,9},
xticklabels={{CW},{},{SC-E1},{},{SC-E2},{},{SC-E3},{},{SC-E4}},
xticklabel style={font=\color{black}, font = \fontsize{20}{1}\selectfont, rotate=0,align=center,yshift=-0.25cm},
ytick={{0},{5},{10},{15},{20},{25},{30},{35},{40},{45},{50},{55},{60},{65},{70},{75},{80},{85},{90},{95},{100}},
yticklabels={{0},{5},{10},{15},{20},{25},{30},{35},{40},{45},{50},{55},{60},{65},{70},{75},{80},{85},{90},{95},{100}},
ymin=50,
ymax=75,
ylabel style={font=\color{black}, font = \fontsize{20}{1}\selectfont},
yticklabel style={font=\color{black}, font = \fontsize{20}{1}\selectfont},
ylabel={Accuracy [\%]},
axis background/.style={fill=white},
xmajorgrids,
ymajorgrids,
legend columns=-1,
legend entries={Long plot title, B, C},
legend style={/tikz/every even column/.append style={column sep=0.5cm}},
legend style={legend cell align=left, align=left, draw=white!15!black,yshift=1.1cm,xshift=-0.25cm, font = \fontsize{18}{1}\selectfont}
]
\addplot[ybar, bar width=0.229, fill=mycolor1, draw=black, area legend] table[row sep=crcr] {%
1	59.2279980945576\\
3	55.5295045849708\\
5	57.3024592116232\\
7	56.4770156008098\\
9	53.6210849112778\\
};
\addlegendentry{$\mathrm{CDR}_{\mathrm{thresh}} = \SI[parse-numbers = false]{-\infty}{\decibel}$}

\addplot[ybar, bar width=0.229, fill=mycolor2, draw=black, area legend] table[row sep=crcr] {%
1	69.6476805216519\\
3	66.458091170725\\
5	65.4517317920072\\
7	62.8988253894382\\
9	63.7404155589471\\
};
\addlegendentry{$\mathrm{CDR}_{\mathrm{thresh}} = \qty{0}{\decibel}$}

\end{axis}
\end{tikzpicture}%
%
	\vskip-1.4cm\caption{Average localization accuracy without (blue) and with (orange) frequency subset selection using an external microphone placed at one of four different positions (SC-E1 -- SC-E4) or not using an external microphone (CW) for the construction of the spatial spectrum.}\vskip-0.25cm
	\label{fig:res}
\end{figure}
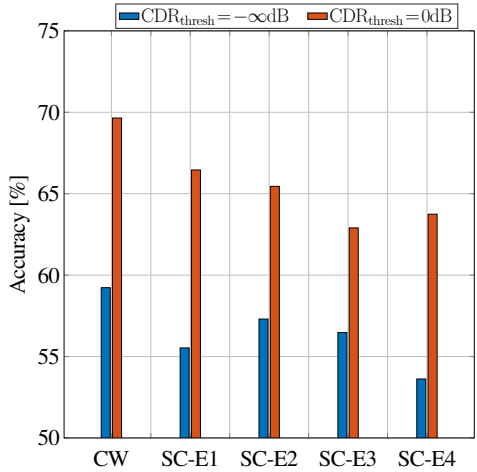
\subsection{Results}\label{subsec_Experiments__res}
Fig. \ref{fig:res} depicts the average localization accuracies that are obtained from the spatial spectrum as in \eqref{eq:spatialSpectrumIwaenc}, denoted by \grqq{}CW\grqq{}, and the accuracies obtained from the spatial spectrum as in \eqref{eq:spatialSpectrum_SC}, denoted by \grqq{}SC-EX\grqq{}, where \grqq{}X\grqq{} stands for one of the four positions of the external microphone. To show the effectiveness of the subset selection, we considered two threshold values, $\mathrm{CDR}_{\mathrm{thresh}} = \SI[parse-numbers = false]{-\infty}{\decibel}$ (corresponding to selecting all frequencies) and $\mathrm{CDR}_{\mathrm{thresh}} = \qty{0}{\decibel}$, shown as blue bars and orange bars, respectively.

First, for every condition a large improvement in the localization accuracy of up to \qty{11}{\percent} due to the frequency subset selection can be observed. This result is in line with the results reported in \cite{Fejgin2022}. Second, considering the spatial spectrum obtained from \eqref{eq:spatialSpectrum_SC}, it can be observed that the position of the external microphone has a minor effect on the estimated DOA, resulting in localization accuracies in the range \qty{62}{\percent} - \qty{66}{\percent} using a threshold value of $\mathrm{CDR}_{\mathrm{thresh}} = \qty{0}{\decibel}$. For the external microphone placed at positions E3 or E4, i.e., close to the loudspeakers playing back the noise, a slightly lower DOA estimation accuracy can be observed when comparing to the external microphone placed at positions E1 or E2. Third, comparing the DOA estimation performance when using the CW method against the SC method for estimating the head-mounted RTF vector, a difference up to around \qty{5}{\percent} - \qty{7}{\percent} can be observed. Thus, the low-complexity SC method yields a comparable DOA estimation performance for multiple speakers as the CW method, which is line with the single speaker DOA estimation results reported in \cite{Fejgin2021}.

\section{Conclusions}
Based on two RTF vector estimation methods, in this paper we compared the DOA estimation performance for multiple speakers for a binaural hearing aid setup exploiting an external microphone or not. We did not restrict the position of the external microphone to be close to the target speaker. Estimating the RTF vector using either the CW method without exploiting the external microphone or using the SC method exploiting the external microphone, we constructed a frequency-averaged Hermitian angle spectrum from which the DOAs of the speakers were estimated as the directions that maximized the spatial spectrum. We evaluated the approach using simulations with recorded two speaker scenarios in acoustic environments with mild reverberation and diffuse-like babble noise scaled to low SNRs for different positions of the external microphone. The results show that using the SC method for the construction of the frequency-averaged Hermitian angle spectrum yields a DOA estimation accuracy (\qty{62}{\percent} - \qty{66}{\percent}) that is comparable to the CW method ($\approx \qty{70}{\percent}$) at a lower computational complexity.

% For bibtex users:
\bibliography{refs.bib}
% For non bibtex users:
%\begin{thebibliography}{citations}
%\bibitem{Author:00}
%E.~Author.
%\newblock The title of the conference paper.
%\newblock In {\em Proc.\ of the European Society on Vibration
%  }, pages 000--111, Chania, Greece, 2018.
%
%\bibitem{Someone:10}
%A.~Someone, B.~Someone, and C.~Someone.
%\newblock The title of the journal paper.
%\newblock {\em Acta Acust united Ac}, A(B):111--222, 2010.
%
%\bibitem{Someone:04}
%X.~Someone and Y.~Someone.
%\newblock {\em The Title of the Book}.
%\newblock S. Hirzel, Stuttgart, Germany, 2012.
%
%\end{thebibliography}
\end{document}